\documentclass{ioppreprint}

\usepackage{bm}
\usepackage{bbm}
\usepackage{amssymb}

\usepackage{graphicx,epstopdf}
\usepackage{xcolor}
\usepackage{amsmath}

\newcommand{\ket}[1]{| #1 \rangle}
\newcommand{\bra}[1]{\langle #1 |}

\newcommand{\SubFig}[2]{\ref{#1}{\color{blue}#2}}

\definecolor{MyGreen}{RGB}{0, 179, 134}
\definecolor{MyRed}{RGB}{255, 102, 102}

\begin{document}

\articletype{} 

\title{Network-Mediated Capacitive Coupling Drives Fast OTOC Saturation in Superconducting Circuits}

\author{Carla Caro Villanova{\color{blue}$^{1,\dagger,\ast}$}\orcid{0009-0002-4562-088X}, and Alan C. Santos{\color{blue}$^{1,\ddagger}$}\orcid{0000-0002-6989-7958}}

\affil{$^1$\href{https://ror.org/009wseg80}{Instituto de Física Fundamental}, \href{https://ror.org/02gfc7t72}{Consejo Superior de Investigaciones Científicas}, Calle Serrano 113b, Madrid 28006, Spain}

\affil{{\color{blue}$^{\ast}$}Author to whom any correspondence should be addressed.}

\email{{\color{blue}$^{\dagger}$}carlacarov@gmail.com, {\color{blue}$^{\ddagger}$}ac\_santos@iff.csic.es}


\begin{abstract}
We investigate the dynamical and spectral consequences of capacitance-network-mediated interactions in superconducting transmon arrays beyond effective nearest-neighbor descriptions. While weak coupling regimes are well captured by effective nearest-neighbor interacting models, we show that increasing capacitive connectivity induces a pronounced departure from this approximation in dynamical observables.
Using Out-of-Time-Ordered Correlators (OTOCs), we demonstrate that such network-mediated couplings significantly accelerate operator scrambling, leading to rapid saturation compared to the nearest-neighbor limit. This dynamical crossover is accompanied by a shift in spectral statistics away from Poissonian behavior toward level repulsion, with the level spacing ratio remaining intermediate between Poisson and Gaussian Orthogonal Ensemble (GOE) limits. This indicates the emergence of partial ergodicity rather than fully developed quantum chaos. As this behavior arises within experimentally realistic regimes of current superconducting transmon devices, identifying when network-mediated couplings qualitatively alter information dynamics is directly relevant for scalable superconducting architectures.
\end{abstract}

\section{Introduction}

Superconducting circuits represent a highly promising and rapidly scaling platform for quantum simulation~\cite{barends:2015,yan:2019,yanay:2020,xiang:2023,shi:2024,rosen:2024,karamlou:2024}, quantum information processing~\cite{barends:2014,wendin:2017,arute:2019,wu:2021,chu:2023}, and quantum communication~\cite{zhong:2021,storz:2023,niu:2023,almanakly:2025}, among other fields~\cite{blais:2021}. Standard architectures typically consist of one-dimensional arrays~\cite{barends:2015,yan:2019} and two-dimensional lattices~\cite{shi:2024,rosen:2024,hu:2026} of superconducting transmon qubits~\cite{koch:2007}, but also other complex structures have been engineered~\cite{wang:2020,yu:2023,hu:2025,javadi-abhari:2025}. In such setups, the system is engineered to enable isolated and highly controllable nearest-neighbor (N.N.) interactions. However, this targeted N.N. connectivity is often accompanied by two additional couplings, namely, \textit{parasitic}~\cite{sung:2021,materise:2023,hu:2023} and \textit{collateral} (or \textit{indirect}) couplings~\cite{yan:2018,yanay:2022,rosario:2023}. 

Parasitic coupling is a next-nearest-neighbor (N.N.N.) interaction between two qubits when the physical distance separating them is not large enough to suppress parasitic capacitances in the device~\cite{materise:2023}. Such parasitic interactions are helpful to speed up two-qubit gates~\cite{santos:2023b}, and also serve as a key ingredient in the mechanism of tunable two-qubit interaction engineering~\cite{yan:2018,hu:2023}. A second kind of interaction is a counter-intuitive capacitive coupling between two arbitrary qubits, $n$ and $m$, induced by the capacitance network between them. This interaction has been named \textit{indirect} interaction~\cite{yan:2018}, or \textit{collateral}~\cite{rosario:2023} interaction, because it exists even in the absence of a direct capacitance connecting the qubits $n$ and $m$~\cite{yan:2018,rosario:2023}. Therefore, such a collateral coupling cannot be explained as a direct spatial parasitic capacitance.

Although the existence of these network-induced interactions has been theoretically identified~\cite{yan:2018,yanay:2022,rosario:2023}, understanding their ultimate impact on information processing remains a relevant open question. In this work, we address this problem for a chain of identical transmon qubits. For the regime of experimentally accessible parameters, we verify the physical consequence of the non-local interactions by tracking the spread of quantum information through the analysis of Out-of-Time-Ordered Correlators (OTOCs)~\cite{shenker:2014,li:2017,garttner:2017,hashimoto:2017,swingle:2018,landsman:2019}. We demonstrate that, even under the effects of collateral coupling, the short-time evolution is strictly described by a N.N. Heisenberg XX model, rendering the dynamics immune to non-local information scrambling. Conversely, these non-local interactions play a crucial role for longer evolution times, leading to the breaking of the system's integrability and a fast saturation of thermal OTOC measurements~\cite{maldacena:2016,lin:2018,lantagne-hurtubise:2020,xu:2024}. However, the system remains in a partially ergodic but non-chaotic regime: unlike recent results reported in quantum annealing~\cite{munoz-arias:2025}, the parameter regime studied here does not exhibit fully developed quantum chaos, as indicated by spectral statistics that remain intermediate between Poisson and GOE behavior. 

\begin{figure}[t!]
	\centering
	\includegraphics[width=1.0\linewidth]{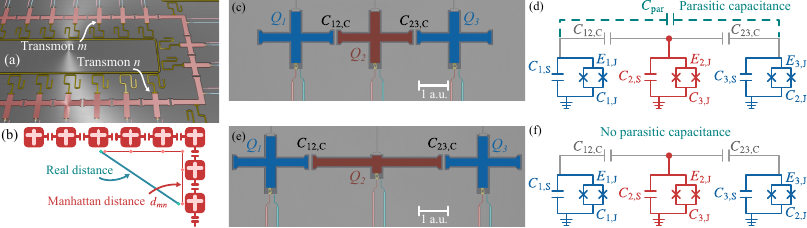}
	\caption{(a) Proposal of a device enabling the experimental implementation of our results, in which a linear chain of transmon qubits can be locally controlled and multi-qubit operations can be performed through nearest-neighbor capacitive interactions. (b) Sketch of the device showing the geometrical meaning of the Euclidean (real) and Manhattan distances. Direct \textit{parasitic} capacitance scales with the Euclidean distance. Conversely, the collateral coupling derived in this work does not follow the Euclidean distance, but rather propagates through the discrete topological links (coupling capacitors), scaling with the ``Manhattan distance'', $d_{mn} = |\vec{n}-\vec{m}|$, across the network graph. (c) Device describing a scenario where a small parasitic coupling between $Q_{1}$ and $Q_{3}$ is present due to their proximity. In this case, the corresponding circuit diagram is shown in (d). The parasitic capacitance $C_{\mathrm{par}}$ is usually much smaller than the direct nearest-neighbor qubit interacting capacitance $C_{n(n+1)}$. (e) System where qubits $Q_{1}$ and $Q_{3}$ are far apart, such that the parasitic coupling vanishes, with the corresponding circuit diagram shown in (f).}
	\label{Fig:Scheme-Plot}
\end{figure}

\section{Nonlocal interactions in transmon qubit arrays} \label{methods}

Our physical setup consists of a 1D array of $N$ superconducting transmon qubits, as depicted in Figs.~\SubFig{Fig:Scheme-Plot}{a} and~\SubFig{Fig:Scheme-Plot}{b}. Each qubit is characterized by the Josephson energy $E_{n,\mathrm{J}}(\Phi_{n}^{\mathrm{ext}})$, tunable via an external magnetic flux $\Phi_{n}^{\mathrm{ext}}$, and by its Josephson capacitance $C_{n,\mathrm{J}}$ and the large shunting capacitance $C_{n,\mathrm{S}}$, which combined define the total capacitance $C_{n,\mathrm{q}} = C_{n,\mathrm{J}} + C_{n,\mathrm{S}}$ for the transmon qubit. The interactions in the system are created through direct transmon-transmon capacitive coupling, $C_{nm}$, between the transmons $m$ and $n$. Therefore, in addition to the nearest-neighbor capacitive coupling, as sketched in Figs.~\SubFig{Fig:Scheme-Plot}{c} and ~\SubFig{Fig:Scheme-Plot}{d}, a weak next-nearest-neighbor interaction is present in the system due to the \textit{parasitic} coupling when the distance between the transmons $n$ and $n+2$ is not large enough to suppress parasitic capacitance, $C_{n(n+2)}^{\mathrm{par}}$ between them. In this case, the circuit diagram is as shown in Fig.~\SubFig{Fig:Scheme-Plot}{d}, where a direct capacitance between the qubits $n$ and $n+2$ needs to be taken into account. On the other hand, and the main focus of this work, a non-trivial scenario is shown in Figs.~\SubFig{Fig:Scheme-Plot}{e} and~\SubFig{Fig:Scheme-Plot}{f}, in which we assume a device characterized by the same capacitance and Josephson energies as in Fig.~\SubFig{Fig:Scheme-Plot}{d}, but now the qubits $n$ and $n+2$ are placed far apart. 

Therefore, in cases where the parasitic capacitance vanishes, the $N$-qubit circuit is an extended version of the diagram shown in Fig.~\SubFig{Fig:Scheme-Plot}{f} for a three-qubit system. The Lagrangian of the system is given by (we use the notation $\dot{\phi} = d\phi(t)/dt$)
\begin{equation}
	\mathcal{L} = \sum_{m,n=1}^N\frac{1}{2}\mathbf{C}_{mn}\dot{\phi}_{m}\dot{\phi}_n + \sum_{m =1}^{N} E_{J,m}(\Phi_{ext}) \cos\left(\frac{2\pi\hat{\phi}_{m}}{\Phi_0}\right) ,
\end{equation}
where $\mathbf{C}$ is the capacitance matrix of the 1D transmons chain given by 
\begin{equation}
	\mathbf{C} =
	\begin{pmatrix}
		C_{1,\mathrm{q}} + C_{12}& -C_{12} & 0  & 0 & \cdots & 0
		\\
		-C_{12} & C_{12} +C_{2,\mathrm{q}} +  C_{23} & -C_{23} & 0  & \cdots & 0
		\\
		0 & -C_{23} &  C_{23} +C_{3,\mathrm{q}} + C_{34}& \ddots & \cdots & \vdots 
		\\
		\vdots & \vdots & \vdots & \cdots & \ddots & -C_{(N-1)N}
		\\
		0 & 0 & 0 & \cdots &  -C_{(N-1)N} & C_{(N-1)N} + C_{N,\mathrm{q}}
	\end{pmatrix} .
	\label{Eq:Capacitance}
\end{equation}

However, even in this scenario, it is surprising that interactions beyond nearest-neighbor qubits can arise in the system. While this interaction is not explicitly shown in the Lagrangian of the system, by applying the Hamiltonian formalism we are able to capture it. Indeed, by following the standard procedure to obtain the Hamiltonian of the system~\cite{vool:2017,rasmussen:2021}, through the Legendre transformation of the Lagrangian, we get (after canonical quantization)
\begin{equation}
	\hat{H} = \frac{1}{2} \sum_{m,n=1}^{N} (\mathbf{C}^{-1})_{mn} \hat{q}_{m} \hat{q}_n - \sum_{m =1}^{N} E_{J,m}(\Phi_{ext}) \cos\left(\frac{2\pi\hat{\phi}_{m}}{\Phi_0}\right) , \label{Eq:1}
\end{equation}
with $\hat{q}_{m}$ and $\hat{\phi}_{m}$ the charge and flux operators for the $m$-th transmon, and $\Phi_0 = \frac{h}{2e}$ the magnetic flux quantum. Now, to highlight the non-local nature of the interactions in the system, it is convenient to separate the Eq.~\eqref{Eq:1} into a self-energy term $\hat{H}_{0}$ and an interaction term $\hat{H}_{\mathrm{int}}$. Using the charge number operator $\hat{n}_{m} = - \hat{q}_{m}/2e$ and defining the local capacitive energy $E_{C,m} = e^2 (\mathbf{C}^{-1})_{mm} / 2$, the self-energy is
\begin{equation}
	\hat{H}_{0} = 4 \sum_{m=1}^{N}E_{C,m}\hat{n}_{m}^{2} - \sum_{m =1}^{N} E_{J,m}(\Phi_{ext}) \cos\left(\frac{2\pi\hat{\phi}_{m}}{\Phi_0}\right) ,
\end{equation}
while the interaction Hamiltonian is correspondingly governed by the off-diagonal elements $E_{C,nm} = e^2 (\mathbf{C}^{-1})_{nm} / 2$, and given by
\begin{equation}
	\hat{H}_{\mathrm{int}} = \sum_{m,n\neq m}^{N} 4E_{C,nm} \hat{n}_{m} \hat{n}_n . \label{Eq:HInt}
\end{equation}

As an immediate consequence of this result, even for a capacitance matrix containing only nearest-neighbor interactions---i.e. a tridiagonal matrix---, in general its inverse is not a tridiagonal matrix, which may lead to transmon-transmon coupling energy $E_{C,nm}$ beyond N.N. interactions. As depicted in Figs.~\SubFig{Fig:Scheme-Plot}{e,f}, the absence of parasitic capacitances $C_{n(n+2)}^{\mathrm{par}}$ in the matrix $\mathbf{C}$ is a clear indication that such indirect couplings are not fully consistent with the introduction of additional parasitic capacitance connecting the qubits $n$ and $m$. Indeed, it must be understood as a collateral effect of a capacitance-network-induced interaction~\cite{rosario:2023}. 

Even for an arbitrary number $N$ of transmons, it is possible to find an exact solution for the spatial profile of the collateral interaction and investigate its impact on the system's dynamical behavior. To this end, first we consider identical transmon-transmon N.N. capacitive coupling $C_{n(n+1)} = C_{\mathrm{C}}$, where the \textit{bulk qubits} of the chain have a capacitance of $C_{n\neq (1,N),q} = C_{\mathrm{q}}$, while the \textit{edge qubit} capacitances are given by $C_{1,q} = C_{N,q} = C_{\mathrm{q}} + C_{\mathrm{C}}$. In this case, the capacitance matrix $\mathbf{C}$ in Eq.~\eqref{Eq:Capacitance} forms a strict symmetric tridiagonal Toeplitz matrix~\cite{dafonseca:2001}
\begin{equation}
	\mathbf{C} = \mathbf{C}_{T} = \begin{pmatrix}
		C_{\mathrm{q}}+ 2C_{\mathrm{C}} & -C_{\mathrm{C}} & 0 & 0 & \dots & 0 \\ 
		-C_{\mathrm{C}} & C_{\mathrm{q}}+2C_{\mathrm{C}} & -C_{\mathrm{C}} & 0 & \dots & 0 \\ 
		0 & -C_{\mathrm{C}} & C_{\mathrm{q}}+2C_{\mathrm{C}} & -C_{\mathrm{C}} & \dots & \vdots \\ 
		\vdots & \vdots & \vdots & \dots & \ddots & -C_{\mathrm{C}} \\ 
		0 & 0 & 0 & \dots & -C_{\mathrm{C}} & C_{\mathrm{q}} + 2C_{\mathrm{C}}
	\end{pmatrix} .
\end{equation}

The exact analytical form for the inverse of the above tridiagonal Toeplitz matrix can be evaluated using a cofactor expansion strategy, where the sub-determinants of the coupling matrix map exactly to Chebyshev polynomials of the second kind, $U_k(z)$ (in Appendix~\ref{Ap:IndenticalShunt} we provide analytical results for the case $C_{1,q} = C_{N,q} = C_{\mathrm{q}}$). For any arbitrary pair of qubits $n$ and $m$ (assuming $m \geq n$ without loss of generality), the inverse matrix element takes the exact form~\cite{dafonseca:2001}
\begin{equation}
	(\mathbf{C}_{\mathrm{T}}^{-1})_{nm} = \frac{1}{C_{\mathrm{C}}} \frac{U_{n-1}(z) U_{N-m}(z)}{U_N(z)}, \quad \mathrm{with~} z = \frac{C_{\mathrm{q}} + 2C_{\mathrm{C}}}{2C_{\mathrm{C}}} . \label{Eq:InvCnm}
\end{equation}

\begin{figure}[t]
	\centering
	\includegraphics[width=1.0\linewidth]{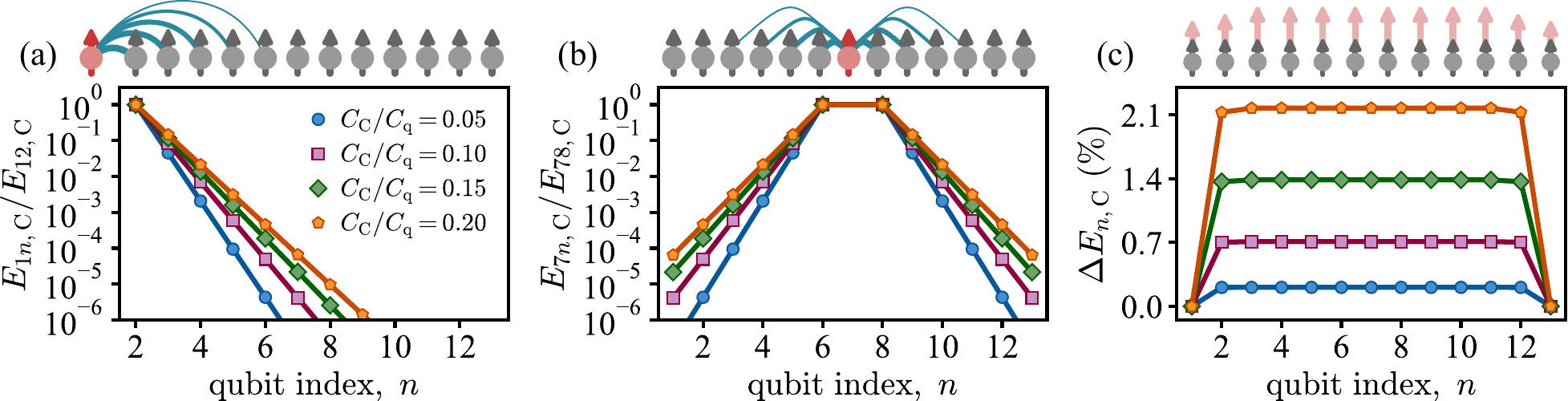}
	\caption{(a,b) Non-local capacitances energies induced by effective capacitance network between the two qubits $m$ and $n$, evaluated across a uniform chain of $N=13$ transmons as a function of the Manhattan distance $d=|n-m|$. In (a) show the interaction between the edge qubit $m=1$ and the $n$-th qubit, and (b) the profile of the non-local interaction with respect to a qubit placed in the center of the chain, namely $m=7$. The exact Chebyshev analytical theory (solid lines) perfectly overlays the raw numerical matrix extraction (symbols). (c) Natural frequency of the qubits showing the clear detuning between the edge qubits ($n = 1$ and $n = N$) with respect to the bulk qubits ($n\neq \{1,N\}$) for different values of the ratio $C_{\mathrm{C}}/C_{\mathrm{q}}$. The simulation utilizes realistic hardware parameters for ground capacitance $C_{\mathrm{q}} = 100$~fF.}
	\label{fig:decay_distance}
\end{figure}

An approximate power law for the non-local capacitive coupling is obtained in the limit where the coupling capacitance is much smaller than the qubit capacitance ($C_C \ll C_{\mathrm{q}}$). In this regime, the argument simplifies to $z \approx C_{\mathrm{q}} / 2C_{\mathrm{C}} \gg 1$. Utilizing the asymptotic limit of the Chebyshev polynomials for large arguments, $U_k(z) \approx (2z)^k$, with $z \approx C_{\mathrm{q}} / 2C_{\mathrm{C}}$, and we recover the effective exchange coupling between any two nodes $n$ and $m$ takes the asymptotic form (again, we use that $m \geq n$)~\cite{yanay:2022,rosario:2023}
\begin{equation}
	(\mathbf{C}_{\mathrm{T}}^{-1})^{z \gg 1}_{nm} \approx \frac{1}{C_{\mathrm{q}}} \left( \frac{C_{\mathrm{q}}}{C_{\mathrm{C}}} \right)^{-|m-n|} , \label{Eq:Asymp}
\end{equation}
which depends on the topological Manhattan distance on the circuit graph as $d = |n-m|$. This demonstrates that, in this limit, the coupling strength is exponentially suppressed with the number of intermediate capacitive links between arbitrary qubits (see Fig.~\ref{fig:decay_distance}), a regime in which the Hamiltonian in Eq.~\eqref{Eq:HInt} is predominantly governed by nearest-neighbor exchange interactions.

However, the range of the non-local interaction can change if the asymptotic limit $C_{\mathrm{q}} / 2C_{\mathrm{C}} \gg 1$ is not satisfied. In Fig.~\ref{fig:decay_distance} we show that the non-local terms will become relevant for larger values of the ratio $C_{\mathrm{q}} / C_{\mathrm{C}}$. In particular, for devices fabricated with capacitive couplings of order of $C_{\mathrm{C}} = 0.2C_{\mathrm{q}}$---which is experimentally feasible~\cite{wu:2024}---, interactions beyond N.N.N. are relevant. For example, in Fig.~\SubFig{fig:decay_distance}{a}, the coupling strength between the first and fourth qubits $E_{14,\mathrm{C}}$ is approximately $2\%$ of the nearest-neighbor interaction $E_{12,\mathrm{C}}$, and the N.N.N. couplings reach values around $15\%$ of $E_{12,\mathrm{C}}$. A similar result is also obtained for qubits in the middle of the chain---see Fig.~\SubFig{fig:decay_distance}{b}. It is worth highlighting that Fig.~\ref{fig:decay_distance} demonstrates accurate agreement between our exact coupling from Eq.~\eqref{Eq:InvCnm} (solid lines) and the exact numerical diagonalization of the capacitance matrix $\mathbf{C}_{\mathrm{T}}$ (symbols).

An additional consequence of the result in Eq.~\eqref{Eq:InvCnm} is an intrinsic ``border effect'' in the local qubit capacitance energy $E_{n,\mathrm{C}}$. In fact, it is possible to show that when $m=n$, we get $(\mathbf{C}_{\mathrm{T}}^{-1})_{nn} \propto U_{n-1}(z) U_{N-n}(z)$, which allows us to conclude that qubits placed at positions $n$ and $N-n$ share the same capacitive energy, as $(\mathbf{C}_{\mathrm{T}}^{-1})_{nn}=(\mathbf{C}_{\mathrm{T}}^{-1})_{(N-n+1)(N-n+1)}$. In particular, this is valid for the edge qubits as $C_{NN}=C_{11}$. As depicted in Fig.~\SubFig{fig:decay_distance}{c}, this leads to a symmetric gradient of frequency along the qubit chain. To quantify how strong the frequency detuning is, we define the quantity $\Delta E_{n,\mathrm{C}} = E_{n,\mathrm{C}}/ E_{1,\mathrm{C}} - 1$, which quantifies the deviation of the energy of the $n$ qubit as a multiple of the energy of the edge qubits $E_{1,\mathrm{C}} = E_{N,\mathrm{C}}$. The behavior of $\Delta E_{n,\mathrm{C}}$ for different ratios can be seen from Fig.~\SubFig{fig:decay_distance}{c}, which makes it clear that the stronger the coupling capacitance, the more detuned the edge qubits are with respect to a given bulk qubit $m$ (around $m = N/2$). 

As we shall see, while relatively small (around $2\%$ of $E_{1,\mathrm{C}}$), the impact of such a border effect in the qubit frequencies requires additional calibration of the parameters of the system for enhanced information control transport.

\section{Quantum information scrambling} \label{results}

Now, we show how information scrambling is affected by the non-local interactions characterized in the previous section. However, we restrict the Hilbert space of our system by writing the Hamiltonian in the two-level system approximation for each transmon, limiting them to the ground and excited states, $\ket{g}$ and $\ket{e}$. In this regime, the system is described by a non-local interacting Heisenberg XX model governed by the Hamiltonian
\begin{equation}
	\hat{H}_{\mathrm{nloc}} = \sum_{m=1}^{N} \hbar\omega_{m} \hat{\sigma}^{+}_{m} \hat{\sigma}^{-}_{m} 
	+ \sum_{m,n>m}^N \hbar g_{nm} \left( \hat{\sigma}^{-}_{n}\hat{\sigma}^{+}_{m} + \hat{\sigma}^{-}_{m}\hat{\sigma}^{+}_{n}\right) , \label{Eq:H_device}
\end{equation}
where $\hat{\sigma}_{m}^{+}$ and $\hat{\sigma}_{m}^{-}$ are raising and lowering spin-$\frac{1}{2}$ Pauli operators for qubit $m$, respectively. The effective qubit frequency is $\hbar \omega_{m} = \sqrt{8E_{C,m}E_{J,m}(\Phi_{ext})} - E_{C,m}$, and $\hbar g_{nm} = 4E_{C,nm}/\sqrt{4\xi_{n}\xi_{m}}$ is the qubit-qubit capacitive coupling, with $\xi_{m} = \sqrt{2 E_{C,m}/ E_{J,m}(\Phi_{ext})}$. To demonstrate this result, we use the transmon approximation to expand the cosine potential up to the fourth order, $\cos\left(\hat{\varphi}\right) \approx \mathbbm{1} - \hat{\varphi}_{m}^2 / 2 +  \hat{\varphi}_{m}^4 / 24$, where $\hat{\varphi}_{m} \equiv 2\pi\hat{\phi}_{m}/\Phi_0$ is the dimensionless flux operator. We then introduce the standard creation and annihilation operators~\cite{rasmussen:2021}
\begin{equation}
	\hat{a}_{m} = \frac{1}{\sqrt{2}} \left(\frac{1}{\sqrt{\xi_{m}}}\hat{\varphi}_{m} - i \sqrt{\xi_{m}}\hat{n}_{m} \right) .
\end{equation}
After these transformations, we apply the Rotating Wave Approximation (RWA) to obtain the Hamiltonian in Eq.~\eqref{Eq:H_device}, which is valid as long as $g_{mn} \ll \omega_{m}, \omega_{n} ~ \forall ~ (m,n)$. Due to the large anharmonicity of the transmons, the two-level system approximation is obtained by imposing the local Hilbert truncation $\hat{a}_m \rightarrow \hat{\sigma}_{m}^{-} = \ket{g}_{m}\bra{e}_{m}$ and $\hat{a}_m^\dagger \rightarrow \hat{\sigma}_{m}^{+} = \ket{e}_{m}\bra{g}_{m}$, such that $\hat{a}_m^\dagger\hat{a}_m^\dagger\hat{a}_m\hat{a}_m = 0$.

To illustrate the impact of the non-local interactions promoted by the collateral couplings, we compare the real device model in Eq.~\eqref{Eq:H_device} with its nearest-neighbor counterpart given by
\begin{equation}
	\hat{H}_{\mathrm{NN}} = \sum_{n=1}^{N} \hbar\omega_{n} \hat{\sigma}^{+}_{n} \hat{\sigma}^{-}_{n} 
	+ \sum_{n=1}^{N-1} \hbar g_{nn+1} \left( \hat{\sigma}^{-}_{n}\hat{\sigma}^{+}_{n+1} + \hat{\sigma}^{-}_{n+1}\hat{\sigma}^{+}_{n}\right) , \label{Eq:NNHamiltonian}
\end{equation}
which is widely used to describe chains of transmon qubits~\cite{rasmussen:2021}, and where $g_{nn+1}$ and $\omega_{n}$ are the same parameters as those used to define the Hamiltonian in Eq.~\eqref{Eq:H_device}. In this model, we assume the resonant condition $\omega_{n} = \omega_{1}$, which is the ideal case for excitation transport. 

\begin{figure}[t]
	\centering
	\includegraphics[width=1.0\linewidth]{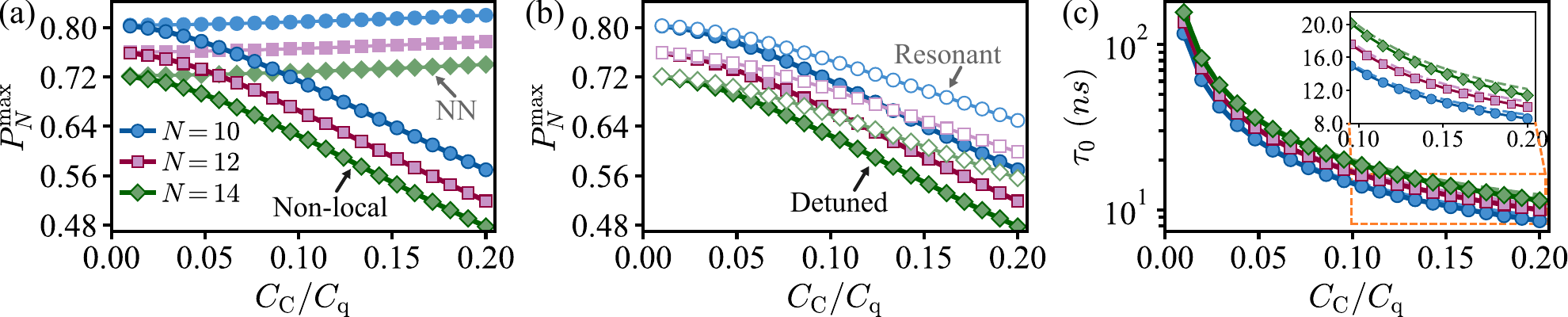}
	\caption{(a) Maximal population transferred from qubit 1 to qubit $N$, for different values of $N$, as a function of the capacitance ratio $C_{\mathrm{C}}/C_{\mathrm{q}}$. While the N.N. Hamiltonian yields a slight enhancement of the transferred population as $C_{\mathrm{C}}/C_{\mathrm{q}}$ increases, the state transfer for the non-local Hamiltonian is progressively degraded. (b) Maximal population reached in qubit $N$ when local qubit flux tunability is used to bring all qubits into resonance, $\omega_{n}=\omega_{1}$, in comparison with the off-resonant case where the edge qubits are detuned. (c) Minimum time required to reach the maximum population shown in panel (a) for the non-local Hamiltonian (solid lines and symbols) and for the N.N. Heisenberg Hamiltonian (dashed lines). The simulations were performed using realistic hardware parameters for a ground capacitance of $C_{\mathrm{q}} = 100$~fF, and a zero-flux Josephson energy of $E_J = 50 \times 10^{-24}$~J.}
	\label{fig:state_transfer}
\end{figure}

Single-qubit information spreading is the most basic application when studying the transport properties of a spin chain~\cite{christandl:2004,gualdi:2008,lewis:2023}. Without loss of generality, we assume the system is initially prepared in the single-excitation state localized at the first qubit, $\ket{\psi(t=0)} = \ket{e}_{1}\ket{g}_{2}\cdots\ket{g}_{N}$. In this protocol, two quantities are particularly relevant: 1) the total population transferred to the last qubit $N$, defined as $P^{\mathrm{max}}_{N} = \max_{t} \bra{\psi(t)} \hat{\sigma}^{+}_{N}\hat{\sigma}^{-}_{N}\ket{\psi(t)}$; and 2) the minimum time $\tau_{0}$ required to transfer the total population $P^{\mathrm{max}}_{N}$ from qubit 1 to qubit $N$ (we select the first maximum value of the population in time). In this scenario, a fundamental question is how non-local capacitive couplings affect these two quantities. By considering different numbers of qubits in the chain, Fig.~\ref{fig:state_transfer} provides information about the maximal population $P^{\mathrm{max}}_{N}$ and minimal time $\tau_{0}$ as a function of the capacitance ratio $C_{\mathrm{C}}/C_{\mathrm{q}}$, indicating a clear disadvantage when operating quantum devices with higher collateral capacitive couplings.

The degradation in the transport properties of the system, observed in Fig.~\SubFig{fig:state_transfer}{a}, is a consequence of two indirect effects driven by the capacitance network. The first effect is the edge qubit detuning shown in Fig.~\SubFig{fig:decay_distance}{c}. In fact, when we impose the resonant condition $\omega_{n} = \omega_{1}$ for the Hamiltonian in Eq.~\eqref{Eq:H_device}, the system is able to transfer more population between the edge qubits, as can be seen in Fig.~\SubFig{fig:state_transfer}{b}. The second effect is related to the qubit-qubit interactions beyond nearest neighbors. For a perfect linear spin chain with only nearest-neighbor interactions, the larger the couplings $g_{nn+1}$, the better the population transfer between the two outer spins (see Fig.~\SubFig{fig:state_transfer}{a}). However, the minimum time required to transfer this maximal population remains almost identical for both models (see Fig.~\SubFig{fig:state_transfer}{c}).

The results discussed so far highlight the negative impact of non-local capacitive interactions in the system, at least for dynamics restricted to the single-excitation manifold (a linear regime of evolution). Conversely, such non-local interactions can provide a positive impact on the system with respect to quantum information scrambling. To demonstrate this result, we go beyond single-excitation dynamics and consider an approach in which multi-excitation processes are taken into account. This can be evaluated through thermal OTOC measurements~\cite{maldacena:2016,lin:2018,lantagne-hurtubise:2020,xu:2024}, which serve as a primary measure of information scrambling for many-body physics in the quantum realm~\cite{swingle:2018}.

\subsection{OTOC Dynamics}

The quantity to be considered here is the dynamical behavior of the infinite-temperature OTOC for two arbitrary observables $\hat{W}$ and $\hat{V}$, given by~\cite{swingle:2017}
\begin{align}
	C(t) = 2 \left[ 1 - \langle \hat{W}^{\dagger}(t)\hat{V}^{\dagger}\hat{W}(t)\hat{V}\rangle_{\infty} \right] ,
\end{align}
where we define $\langle \hat{A} \rangle_{\infty} = \mathrm{tr}(\hat{A}\hat{\rho}_{\infty})$, with $\hat{\rho}_{\infty} = \mathbbm{1}/2^{N}$ the thermal density matrix of the system at infinite temperature—that is, $\hat{\rho}_{\infty} = \mathcal{Z}^{-1} e^{-\beta \hat{H}}$ in the limit $\beta \to 0$, with $\mathcal{Z}=\mathrm{tr}\big(e^{-\beta \hat{H}}\big)$ the partition function. In this way, we can understand the global information scrambling and transport behavior from a state-independent quantity. The time-dependent behavior of the operator $\hat{W}(t)$ is computed from the evolution operator $\hat{U}(t)$ as $\hat{W}(t) = \hat{U}^{\dagger}(t) \hat{W}\hat{U}(t)$. Throughout our work, we consider the OTOC dynamics for two observables of the edge qubits by setting $\hat{W} = \hat{\sigma}_{1}^{z}$ and $\hat{V} = \hat{\sigma}_{N}^{z}$.

\begin{figure}[t]
	\centering
	\includegraphics[width=1.0\linewidth]{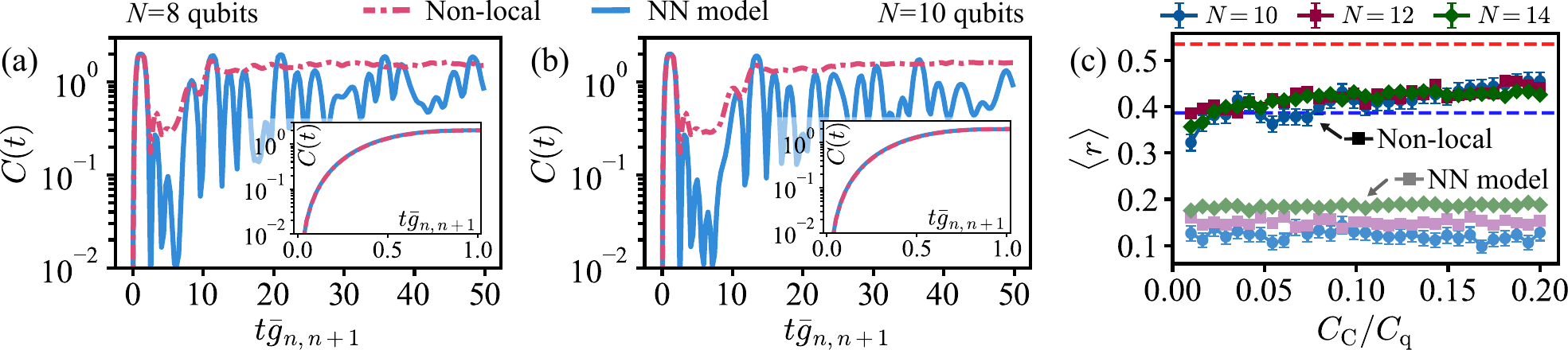}
	\caption{Dynamics of OTOCs for systems (a) with $N=8$ and (b) $N=10$ transmons as a function of the normalized time $t \bar{g}_{nn+1}$, where $\bar{g}_{nn+1}$ is the average N.N. coupling defined as $\bar{g}_{nn+1}=\sum_{n=1}^{N-1}g_{nn+1}/(N-1)$. The solid line denotes the expected behavior of the N.N. Hamiltonian in Eq.~\eqref{Eq:NNHamiltonian}, while the dot-dashed line represents the dynamical OTOC considering the full Hamiltonian in Eq.~\eqref{Eq:H_device}. (c) Mean level spacing ratio for the non-local Hamiltonian, Eq.~\eqref{Eq:H_device}, and its N.N. counterpart, Eq.~\eqref{Eq:NNHamiltonian}, as a function of the capacitance ratio for systems of different sizes. The blue and red dashed lines indicate the level-spacing ratio corresponding to the Poissonian distribution, $\langle r \rangle_{\mathrm{P}}$, and the Gaussian Orthogonal Ensemble (GOE), $\langle r \rangle_{\mathrm{GOE}}$, respectively. The hardware parameters are the same as in Fig.~\ref{fig:state_transfer}.}
	\label{fig:corr_growth}
\end{figure}

We show, in Figs.~\SubFig{fig:corr_growth}{a} and~\SubFig{fig:corr_growth}{b}, the dynamics of the OTOC for transmon chains with $N=8$ and $N=10$ qubits, respectively. We consider two values of the ratio $C_{\mathrm{C}}/C_{\mathrm{q}}$, namely $C_{\mathrm{C}}/C_{\mathrm{q}} = 0.02$, which approximately behaves like a nearest-neighbor (N.N.) model, and $C_{\mathrm{C}}/C_{\mathrm{q}} = 0.15$. As expected, the N.N. model shows a behavior consistent with coherent information spreading over the chain ($\mathrm{max}_{t}\;C(t) = 2$) with multiple reflections across the finite chain ($C(t) \sim 10^{-1}$). However, as a primary result from this dynamics, the identical initial increase of $C(t)$ for the non-local and N.N. Hamiltonians demonstrates that both models are dominated by N.N. interactions at early times (see inset plots in Figs.~\SubFig{fig:corr_growth}{a} and~\SubFig{fig:corr_growth}{b}). 

On the other hand, the OTOC dynamics for the non-local Hamiltonian follows a drastically different behavior when the evolution time becomes large enough, as in this regime the non-local terms significantly contribute to the dynamics. Consequently, we observe that the OTOC exhibits a smoother and faster approach to saturation at a non-zero value of the correlation $C(t)$, reflecting the additional long-range interaction channels that enhance operator scrambling. While such fast saturation is often observed in chaotic systems, the existence of scrambling does not necessarily imply a chaotic regime~\cite{xu:2020,dowling:2023}. Indeed, we will now demonstrate that in this particular case, there is no evidence of a chaotic regime for the device parameters considered.

To this end, we follow the standard diagnostic for quantum chaos and for detecting integrability breaking in the transmon network by analyzing the spectral statistics of the system’s Hamiltonian~\cite{oganesyan:2007}. For a system with eigenenergies $E_n$, the distribution of level spacings $s_n = E_{n+1} - E_n$ distinguishes between integrable (Poissonian) and chaotic (Wigner-Dyson) dynamics. This is quantified by the mean level spacing ratio $\langle r \rangle = \langle \text{min}(s_{n+1},s_n)/\text{max}(s_{n+1},s_n)\rangle$, which yields $\langle r \rangle_{\mathrm{P}} \approx 0.386$ for Poisson and $\langle r \rangle_{\mathrm{GOE}} \approx 0.535$ for the Gaussian Orthogonal Ensemble (GOE) distributions~\cite{oganesyan:2007,atas:2013}.

By imposing only nearest-neighbor interactions, the mean ratio remains below the Poissonian limit ($\langle r \rangle_{\mathrm{NN}} < \langle r \rangle_{\mathrm{P}}$) across the entire range of coupling capacitances from $C_\mathrm{C} \ll C_\mathrm{q}$ up to $C_\mathrm{C}/C_\mathrm{q} = 0.2$, for systems with $N=10$, $N=12$, and $N=14$ qubits (see Fig.~\SubFig{fig:corr_growth}{c}). Therefore, in all these cases, the system behaves as expected and obeys Poissonian statistics characteristic of integrable models. However, this integrability can be easily broken when non-local capacitive interactions are assumed. Indeed, for the real device case with interactions beyond the N.N. capacitive coupling, the non-local interactions induce a breakdown of integrability for values of $C_\mathrm{C}/C_\mathrm{q}$ accessible in current transmon~\cite{zotova:2024,wu:2024,li:2025} and fluxonium-transmon hybrid devices~\cite{ding:2023}. 

This is witnessed by values of $\langle r \rangle_{\mathrm{nLoc}} \geq \langle r \rangle_{\mathrm{P}}$ reached as the ratio $C_\mathrm{C}/C_\mathrm{q}$ increases, leading to a breakdown of the strict one-dimensional tight-binding structure of the Hamiltonian. However, for the range of capacitance $C_\mathrm{C}$ considered in our model—required to satisfy the RWA and the two-level approximation for transmon qubits—the statistics do not approach $\langle r \rangle_{\mathrm{GOE}}$. The fact that the system remains below the level of repulsion associated with the GOE confirms that the network-induced collateral couplings, despite their non-local nature and their influence in accelerating quantum information scrambling, do not introduce sufficient stochasticity to drive the system into a chaotic regime (red dashed line in Fig.~\SubFig{fig:corr_growth}{c}).

\section{Conclusions}

In this work, we revisited the characterization of network-induced capacitive interactions in superconducting transmon arrays~\cite{yan:2018,yanay:2022,rosario:2023}, providing an exact analytical treatment in terms of Chebyshev polynomials. This formulation provides a compact description of the full capacitance network and enables a controlled analysis of interaction terms beyond the nearest-neighbor approximation. Such non-local couplings naturally arise within experimentally relevant parameter regimes and can impact controllability in both analog and digital quantum information processing. In particular, increased capacitive connectivity introduces additional interaction channels that may complicate Hamiltonian engineering in analog quantum simulators. At the same time, these non-local interactions have pronounced consequences for information dynamics. The results reported here provide clear signatures of the potential emergence of quantum-chaotic behavior induced by network-mediated couplings. However, within the range of parameters and system sizes explored, we do not find evidence of a fully developed chaotic regime. 

Our results suggest that enhanced capacitive couplings can significantly modify dynamics without necessarily leading to uncontrolled behavior, opening the possibility of leveraging such effects while maintaining a degree of controllability in superconducting circuit architectures. For instance, the non-local interactions can be used for the fast creation of non-local correlations~\cite{touil:2020,touil:2024}. Furthermore, rapid OTOC saturation can accelerate information-scrambling-based metrology protocols~\cite{kobrin:2024,montenegro:2025,liu:2026,sharma:2026}, which have recently been implemented in superconducting devices~\cite{ge:2025,hu:2026c}.

\ack{CCV acknowledges financial support from the Consejo Superior de Investigaciones Científicas (CSIC) through the Becas JAE Intro ICU 2025 program (Ref. JAEICU\_25\_04998). ACS acknowledges financial support from the Comunidad de Madrid through the program Talento 2024 `César Nombela', Grant No. 2024-T1/COM-31530 (Project SWiQL).}  





\appendix

\section{Capacitive coupling for identical shunting capacitances $C_{n,\mathrm{q}} = C_{\mathrm{q}}$} \label{Ap:IndenticalShunt}

We start from Eq.~\eqref{Eq:Capacitance} for a chain of $N$ qubits and assume identical qubit capacitances, $C_{n,\mathrm{q}} = C_{\mathrm{q}}$. In this case, the capacitance matrix $\mathbf{C}$ is tridiagonal with nearest-neighbor couplings $C_{i,i\pm1} = -C_C$. It is convenient to factor out $C_C$ and introduce the dimensionless parameter $z = (C_{\mathrm{q}} + 2C_C)/2C_C$, such that $\mathbf{C} = C_C \mathbf{M}$, where
\begin{equation}
	\mathbf{M} =
	\begin{pmatrix}
		2z-1 & -1 & 0 & \cdots & 0 \\
		-1 & 2z & -1 & \cdots & 0 \\
		0 & -1 & 2z & \ddots & \vdots \\
		\vdots & \ddots & \ddots & \ddots & -1 \\
		0 & \cdots & 0 & -1 & 2z-1
	\end{pmatrix}.
\end{equation}

To compute the inverse matrix elements, we use a standard result for symmetric tridiagonal matrices. For $n \le m$, the elements are given by~\cite{mallik:2001}
\begin{equation}
	(\mathbf{C}^{-1})_{nm} = \frac{1}{C_C} \frac{\theta_{n-1}\,\phi_{m+1}}{\Delta_N},
	\label{Ap:Eq:Cinv_general}
\end{equation}
where $\Delta_N = \det(\mathbf{M})$, $\theta_k$ is the determinant of the leading principal submatrix $\mathbf{\Theta}_k$ of order $k$,
\begin{equation}
	\mathbf{\Theta}_k =
	\begin{pmatrix}
		2z-1 & -1 & 0 & \cdots & 0 \\
		-1 & 2z & -1 & \cdots & 0 \\
		0 & -1 & 2z & \ddots & \vdots \\
		\vdots & \ddots & \ddots & \ddots & -1 \\
		0 & \cdots & 0 & -1 & 2z
	\end{pmatrix}_{k \times k},
\end{equation}
and $\phi_k$ is the determinant of the trailing principal submatrix $\mathbf{\Phi}_k$,
\begin{equation}
	\mathbf{\Phi}_{N-m} =
	\begin{pmatrix}
		2z & -1 & 0 & \cdots & 0 \\
		-1 & 2z & -1 & \cdots & 0 \\
		0 & -1 & 2z & \ddots & \vdots \\
		\vdots & \ddots & \ddots & \ddots & -1 \\
		0 & \cdots & 0 & -1 & 2z-1
	\end{pmatrix}_{(N-m)\times(N-m)}.
\end{equation}

To find an explicit expression, we first evaluate $\theta_k = \det(\mathbf{\Theta}_k)$. Expanding along the last row yields the recurrence relation
\begin{equation}
	\theta_k = 2z\,\theta_{k-1} - \theta_{k-2},
\end{equation}
with initial conditions $\theta_0 = 1$ and $\theta_1 = 2z-1$. This recurrence coincides with that of the Chebyshev polynomials of the second kind $U_k(z)$, which satisfy $U_k(z) = 2z\,U_{k-1}(z) - U_{k-2}(z)$. Matching the initial conditions gives~\cite{mallik:2001, dafonseca:2007}
\begin{equation}
	\theta_k = U_k(z) - U_{k-1}(z).
\end{equation}

A completely analogous recurrence is obtained for $\phi_k = \det(\mathbf{\Phi}_k)$. Although $\mathbf{\Phi}_k$ differs from $\mathbf{\Theta}_k$ in the position of the boundary term $2z-1$, the two matrices are related by a reversal of the ordering of rows and columns. Such a permutation does not change the determinant; therefore, $\phi_k$ satisfies the same recurrence with the same initial conditions, leading to $\phi_k = \theta_k$.

Finally, the full determinant $\Delta_N = \det(\mathbf{M})$ is computed by expanding along the first row of $\mathbf{M}$:
\begin{equation}
	\Delta_N = (2z-1)\theta_{N-1} - \theta_{N-2}.
\end{equation}
Using the recurrence relations and the identity for $\theta_k$, this simplifies to
\begin{equation}
	\Delta_N = U_N(z) - 2U_{N-1}(z) + U_{N-2}(z).
\end{equation}

Substituting these results into Eq.~\eqref{Ap:Eq:Cinv_general}, we obtain the analytical expression for the inverse matrix elements for $n \le m$:
\begin{equation}
	(\mathbf{C}^{-1})_{nm} = \frac{1}{C_C} \frac{[U_{n-1}(z) - U_{n-2}(z)][U_{N-m}(z) - U_{N-m-1}(z)]}{U_N(z) - 2U_{N-1}(z) + U_{N-2}(z)}.
\end{equation}


\end{document}